\documentclass[journal, 10pt]{IEEEtran}
\usepackage{graphicx}
\usepackage{epsfig}
\usepackage{algorithm}
\usepackage{algorithmic}
\usepackage{subfigure}
\usepackage{ifmtarg}
\usepackage[font=small,labelsep=none]{caption}
\usepackage{footnote}
\usepackage{cite}
\usepackage{amssymb}
\usepackage{amsthm}
\usepackage[fleqn]{amsmath}
\usepackage{amsmath}
\usepackage{amsfonts}
\usepackage{color}
\usepackage{bbm}
\usepackage{cite}
\usepackage{epstopdf}
\usepackage[table,xcdraw]{xcolor}
\usepackage{flushend}

\newtheorem{theorem}{\textbf{\text{Theorem}}}

\usepackage{flushend}

\begin{document}
\title{In-Band Full-Duplex Communications for Cellular Networks with Partial Uplink/Downlink Overlap}
\author{
\IEEEauthorblockN{\large  Ahmad AlAmmouri, Hesham ElSawy, Osama Amin, and Mohamed-Slim Alouini\\
\IEEEauthorblockA{\small Computer, Electrical, and Mathematical Sciences and Engineering (CEMSE) Divison,\\ King Abdullah University of Science and Technology (KAUST), Thuwal, Makkah Province, Saudi Arabia,\\  Email: \{ahmad.alammouri, hesham.elsawy, osama.amin, slim.alouini\}@kaust.edu.sa}
\thanks{The work of M. -S. Alouini was supported by the Qatar National Research Fund (a member of Qatar Foundation) under NPRP Grant NPRP 5-250-2-087. The statements made herein are solely the responsibility of the authors. }}\vspace{-1.0cm}}

\maketitle
\thispagestyle{empty}
\pagestyle{empty}

\begin{abstract}

In-band full-duplex (FD) communications have been optimistically promoted to improve the spectrum utilization in cellular networks. However, the explicit impact of spatial interference, imposed by FD communications, on uplink and downlink transmissions has been overlooked in the literature. This paper presents an extensive study of the explicit effect of FD communications on the uplink and downlink performances. For the sake of rigorous analysis, we develop a tractable framework  based on stochastic geometry toolset. The developed model accounts for uplink truncated channel inversion power control in FD cellular networks. The study shows that FD communications improve the downlink throughput at the expense of significant degradation in the uplink throughput. Therefore, we propose a novel fine-grained duplexing scheme, denoted as $\alpha$-duplex scheme, which allows a partial overlap between uplink and downlink frequency bands. To this end, we show that the amount of the overlap can be optimized via adjusting $\alpha$ to achieve a certain design objective.


\end{abstract}


\section{Introduction}\label{Introduction}

Due to the overwhelming effect of self-interference (SI), wireless transmission and reception are always separated in time, denoted as time division duplexing (TDD), or in frequency, denoted as frequency division duplexing (FDD). Recent advances in transceiver design tend to make SI cancellation viable and alleviate the necessity of such time/frequency separation \cite{FD1, FD_mag}. That is; SI cancellation techniques enable transceivers to achieve acceptable isolation between transmit and receive circuitries while transmitting and receiving on the same time-frequency resource block. It is argued that exploiting the entire bandwidth for FDD, or time in TDD, systems for transmission and reception, denoted as in-band FD communications, can double the spectral efficiency and improve the network capacity\cite{FD1}. This argument makes the in-band FD  a good candidate technology for cellular operators to cope with the challenging performance metrics defined for 5G cellular network \cite{FD_mag}.


SI cancelation techniques enable transceivers to cancel their own SI only but not the interference coming from other sources reusing the same frequency over the spatial domain. Therefore, in multi-access network setup with spatial frequency reuse, the additional interference imposed by FD communications may diminish the harvested performance gain.  Hence, the study of the aggregate (i.e., spatial) interference is essential to characterize the FD performance in large-scale wireless networks. In this regards, stochastic geometry is a powerful tool for such performance characterization \cite{survey_H, FD_LSN, FD_cellular2, Que}. For instance, in the context of ad-hoc networks, \cite{FD_LSN} shows that FD communications can improve the overall throughput in spite of the increased aggregate interference level. Further, in the context of cellular networks, \cite{FD_cellular2} and \cite{Que} shows that FD communications effectively improve the downlink performance. However, the explicit effect of FD communications on uplink performance has been overlooked. While the superiority of downlink over uplink performance has been proved in the case of half-duplex (HD) operation \cite{uplink_h, uplink_2}, the vulnerability of uplink in the case of FD operation has never been studied.

In this paper, we present a comprehensive study of the explicit effect of FD operation on the uplink and downlink scenarios. We show that the uplink is sensitive to the downlink interference, denoted in the sequel as cross-mode interference, and reveal that the harvested FD gain in the downlink may come at the expense of a significant degradation in the uplink performance. Therefore, we propose a novel fine-grained duplexing scheme, denoted as $\alpha$-duplex, which allows a partial overlap between uplink and downlink spectrum. The amount of the overlap is controlled via a design parameter $\alpha$ to balance the tradeoff between the uplink and downlink performance. It is worth mentioning that the proposed $\alpha$-duplex scheme captures the FD and traditional HD schemes as special cases. Specifically, setting $\alpha$ to one enforces FD communications while setting $\alpha$ to zero maintains the conventional HD communications. For the proposed study, we exploit stochastic geometry\footnote{Stochastic geometry has been widely applied in the cellular networks domain as they exhibit random patterns rather than idealized grids \cite{tractable_app, martin_ppp}.} to develop a tractable model, that accounts for the per-user power control and the limited transmission power of user equipments (UEs), for FD based cellular networks.

The rest of the paper is organized as follows: in Section II, we present the system model, assumptions, and methodology of the analysis. In Section III, we analyze the performance of the $\alpha$-duplex system. Numerical and simulation results are presented in Section IV before presenting the conclusion in Section V.

\section{System Model And Methodology}\label{System Model}
\subsection{Network Model}

For simplicity and tractability, we consider a bi-dimensional  single-tier cellular network in which the locations of the base stations (BSs) are modeled as a homogeneous Poisson point process (PPP) $\Psi=\{x_i,i=1,2,3,.... \}$ with intensity $\lambda$, where $x_i \in \mathbb{R}^2$ denotes the location of the $i^{th}$ BS \cite{uplink_h}. The PPP assumption for cellular networks is justified by experimental studies in \cite{tractable_app, martin_ppp} and a theoretical study in \cite{valid}. The locations of the UEs are modeled via an independent PPP  $\Psi_u$ with intensity $\lambda_u \gg \lambda$ such that each BS will always have a user to serve. A general power-law path-loss model is assumed so the signal power decays at a rate $r^{-\eta}$ with the distance $r$, where $\eta>2$ is the path-loss exponent. In addition to the path-loss attenuation, transmitted signals in uplink and downlink experience Rayleigh fading channels with independent and identically distributed (i.i.d) channel gains, and hence, the channel power gains are exponentially distributed random variables and with unity means.

All BSs have the same transmit power $P_b$ in the downlink and UEs have a maximum transmit power $P^{(m)}_u$ and employ a truncated channel inversion power control scheme in which each UE compensates for the path-loss to maintain a target average power level of $\rho$ at the serving BS. Users that cannot maintain the required power level of $\rho$ do not transmit and go to outage\footnote{Truncation outage and the effect of allowing UEs in truncation to transmit with the maximum transmit power has been studied in \cite{uplink_h, uplinkConf} and are out of the scope of the current paper.}. In the analysis, we consider only UEs that satisfy the channel inversion power control and maintain an average power $\rho$ at their serving BSs, which we denote as active UEs. In this case , following \cite{uplink_h}, it is easy to show that the distribution of the distance between an active UE and its serving BS is,

\small
 \begin{align}\label{equ:rDis}
\!\!\!\!\!\!\!\!\!\!\!\!f_R(r)=\frac{2 \pi \lambda r \exp\left(- \pi \lambda r^2 \right) }{1-\exp\left(- \pi \lambda \left(\frac{P^{(m)}_u}{\rho}\right)^{\frac{2}{\eta}} \right)}\mathbbm{1}_{\left\{0 \le r\le \left(\frac{P^{(m)}_u}{\rho}\right)^{\frac{1}{\eta}}\right\}}(r),
\end{align}\normalsize
where $\mathbbm{1}_{\{.\}}(.)$ is the indicator function that equals to one if the condition$\{.\}$ is satisfied and zero otherwise. The moments of the uplink transmit power for an active UE is expressed as~\cite{uplink_h},

\small
\begin{equation}
    \mathbb{E}\left[{P_u}^{\alpha} \right]=\frac{\rho^{\alpha} \gamma\left(\frac{\alpha \eta}{2}+1, \pi \lambda \left(\frac{P^{(m)}_u}{\rho}\right)^{\frac{2}{\eta}} \right)}{(\pi\lambda)^{\frac{\alpha \eta}{2}}\left( 1-\exp\left(- \pi \lambda \left(\frac{P^{(m)}_u}{\rho}\right)^{\frac{2}{\eta}} \right)\right)}.
    \label{equ:powerD}
\end{equation}
\normalsize
\subsection{$\alpha$-Duplex Model}

\begin{figure}[t]
\centerline{\includegraphics[width=  2.45 in]{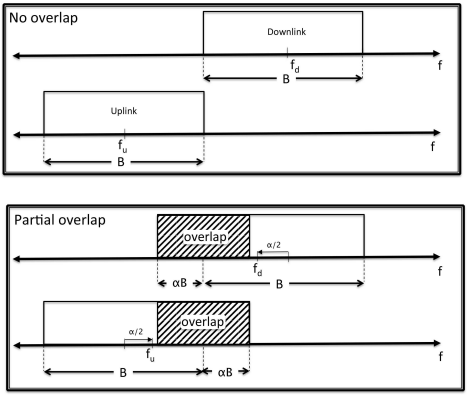}}\caption{\,A schematic diagram of the proposed $\alpha$-duplex scheme for $B_u = B_d  = B$ .}
\label{fig:proposed}
\end{figure}

Universal frequency reuse is adopted with no intra-cell interference. Hence, only one UE per BS is active at a given frequency channel. Without loss of generality, we focus on the case in which downlink and uplink occupy two non-overlapping adjacent null-to-null bands of $B_d$ and $B_u$, respectively, in which the carrier frequencies $f_d > f_u$. SI cancelation capability is exploited to increase the spectral efficiency and allow partial overlap of $2 \alpha B$, where $B = \min(B_u,B_d)$,  between uplink and downlink transmissions, as shown in Fig.~\ref{fig:proposed}. That is, an uplink UE is allowed to access a total of $B_u + \alpha B$ where $\alpha B$ is consumed from the downlink band. Similarly, a downlink BS is allowed to access a total of $B_d + \alpha B$ where $\alpha B$ is consumed from the uplink band. It is worth noting that, since the transmission bandwidth (BW) is a function of $\alpha$, the center frequencies for the uplink and downlink are also function of $\alpha$. According to our system model, the difference between the downlink center frequency ($f_d$) and the uplink center frequency ($f_u$) is given~by,
\begin{align}
    f_d-f_u=\frac{B_u+B_d}{2}-\alpha B.
\end{align}

In this paper, we conduct the analysis on a test link in which the receiver is assumed to exist at the origin. According to
Slivnyak's theorem [8], there is no loss of generality in this assumption. As a result of the uplink/downlink spectrum overlap, the test receiver will experience interference from downlink BSs as well as uplink UEs. Note that, despite the initial assumption of independent PPPs for the UEs and BSs, the set of interfering UEs are not a PPP, due to the no intra-cell interference condition, and the locations of the interfering BSs and UEs are correlated, due to the association technique. However, to maintain the model tractability, we assume that the set of interfering UEs constitutes a PPP $\Phi$, with intensity $\lambda$ which is independent of the set interfering BSs. Note that the intensity of the interfering UEs is selected to be equal to the intensity of the BSs ($\lambda$) because the system does not allow two UE served by the same BS to use the same channel. It is important to highlight that the PPP assumption for the interfering UEs and the independent PPP assumption for interfering BSs and UEs are mandatory for the model tractability and have been used before in \cite{uplink_2, uplink_h, Que, FD_cellular2}. These assumptions ignore the mutual correlation between the interfering sources, however, the correlation between the interfering sources and the test receiver is captured through the proper calculation for the interference exclusion region enforced by association and/or  uplink power control. The accuracy of these approximations is validated via independent simulation in Section~\ref{Results}.

\subsection{Base-band Signal Representation}
\normalsize
The data at the test transmitter, which is a UE in the uplink and BS in the downlink, is mapped to a general bi-dimensional and symmetric constellation with unit energy. For signal transmission, the BSs use the pulse shape $s_d(t) \overset{FT}{\longleftrightarrow} S_d(f)$ in the downlink, and the UEs use the pulse shape $s_u(t) \overset{FT}{\longleftrightarrow} S_u(f)$ in the uplink, where $FT$ denotes the Fourier transform (FT). The used pulse shapes in the uplink and downlink can be  similar or different. At the receiver side, the signal is passed through a matched filter $H(f)$ before taking a decision at the decoder, where $\int_{\mathrm{band}} |H(f)|^2df=1$, in which $H(f)=S_u^{*}(f)$ for the uplink, $H(f)=S_d^{*}(f)$ for the downlink, and $S^{*}$ denotes the conjugate of $S$. Taking the uplink as an example, the received uplink baseband signal at the input of the matched filter of test BS can be expressed~as

\small
\begin{align}
\!\!\!\!\! \!\!\!\!\!\! y_u(t)=  A \sqrt{\rho h_o} s_u(t) + \sum_{k \in \Psi} i_{k}^{d}(t)+ \sum_{j \in \Phi}i_{j}^{u}(t) + i_s(t) + n(t),
\label{base_band1}
\end{align}
\normalsize
where, $\rho$ is the average received power at the BS due to the channel inversion power control, $A$ represents the complex symbol of interest, $h_o$ is the intended channel power gain, $s_u(t)$ is the pulse shape, $ \sum_{k \in \Psi} i_{k}^{d}(t)$  is the aggregate downlink (i.e., cross-mode) interference, $\sum_{j \in \Phi}i_{j}^{u}(t) $ is the aggregate uplink interference,  $i_s$  is the self-interference, and $n(t)$ is a white complex Gaussian noise with zero mean and two-sided power spectral density $N_o/2$.

To facilitate the analysis, we abstract symbols from interfering sources via Gaussian codebooks as in \cite{Hamdi, Gaus_approx1}. The accuracy of the Gaussian codebook approximation for interfering symbols from several constellation types have been verified in \cite{Gaus_approx, Gaus_approx1}. In this case, the interference terms can be expressed as,
\begin{align}
&i_{k}^{d}(t)=  \Gamma_{d_{k}} s_d(t) \sqrt{P_{b_k} h_k r_k^{-\eta}} \exp \left( j 2 \pi (f_d-f_u)t \right), \notag \\
&i_{j}^{u}(t)=  \Gamma_{u_{j}} s_u(t) \sqrt{P_{u_j} h_j r_j^{-\eta}},  \notag \\
&i_{s}(t)=  \Gamma_{s} \sqrt{\beta P_b} s_d(t)\exp \left( j 2 \pi (f_d-f_u)t \right),
\end{align}
where, $\Gamma_{d_{k}}$, $\Gamma_{u_{j}}$, and $\Gamma_s$ are independent complex Gaussian random variables with zero mean and unit variance that represent the interfering symbol from, respectively, the $k^{th}$ interfering BS, the $j^{th}$ interfering UE, and the self-interference. $h_k$'s and $h_j$'s are the channel power gains, $ r_k$ and $r_j$ are the distances between the tagged BS and the $k^{th}$ interfering BS and the $j^{th}$ interfering UE, respectively. $P_{u_j}$ is the transmitted power of the $j^{th}$ interfering UE and $P_{b_k}$ is the transmitted power of the $k^{th}$ interfering BS, since we assume that all BSs transmit by a  fixed power, we will drop the index $k$ and donate BS transmit power by $P_{b}$. $\beta$ represents the self-interference attenuation, which is set to zero if perfect SI is achieved. A representation similar to \eqref{base_band1} can be written for the downlink baseband signal received at the input of the matched filter of the test UE.
\subsection{Methodology of Analysis}
The main performance metrics of interest are the spatially averaged per link bit error rate (BER) and throughput. Assuming a maximum likelihood detector and coherent modulation schemes, the BER in mostly in the form  $\textbf{BER} \approx \omega_1 \; \text{erfc}(\sqrt{ \omega_2 \;  \rm SINR})$, where $\omega_1$ and $\omega_2$ are modulation specific constants and the SINR inside the error function is the deterministic signal power averaged over Gaussian interference plus Gaussian noise, the values of $\omega_1$ and $\omega_2$ are given in \cite[Table 6.1]{Goldsmith} for different modulation schemes. Also, the throughput will be proportional to the BW multiplied by the BER as $T = \log_2{( M)} \rm BW (1- \textbf{BER})$, where $M$ is the number of symbols. It is important to highlight that we  neither have a deterministic signal power nor a Gaussian interference in the depicted system model. Hence, to use the aforementioned expressions, we have to condition on the received signal power and express the interference as a conditionally Gaussian random variable in order to compute the conditional BER via the aforementioned expressions. Then, the actual BER is calculated by an additional averaging step as $\mathbb{E}[\omega_1 \text{erfc}(\sqrt{\omega_2 SINR})]$.
\section{Performance Analysis}
The first step in the analysis is to express the base-band signal after the matched filter. In the uplink case, the base-band signal after the matched filter at $t=t_o$ can be expressed~as

\begin{figure}[!ht]
\centerline{\includegraphics[width=  2.95in]{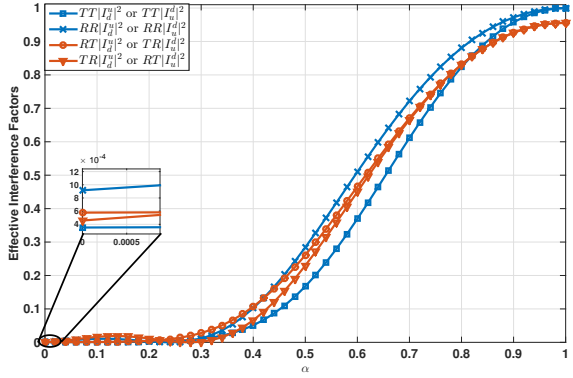}}\caption{\,:$|I^u_d|^2$ $|I^d_u|^2$ vs $\alpha$ for different pulse shapes, where T stands for triangle pulse shape and R for rectangle pulse shape. The first letter represents the pulse shape for the downlink, while the latter is for the uplink.}
\label{fig:EI}
\end{figure}

\small
\begin{align}
&\!\!\!\!\!\!\!\!\!\!\!\!\!y_u(t_o)=y_u(t) * s_u(t-t_o)|_{t=t_o}\int\nolimits_{-\frac{B_u+\alpha B}{2}}^{\frac{B_u+\alpha B}{2}} Y_u(f) S^{*}_u(t)df \notag \\
  &\!\!\!\!\!\!\!\!\!\!\!\!\!=A\sqrt{\rho h_o} + \sum_{k \in \Psi} \Gamma_{d_{k}} \mathcal{I}^{(u)}_{d}(\alpha) \sqrt{P_b h_k r_k^{-\eta}}  \notag \\
 &\!\!\!\!\!\!\!\!\!\!\!\!\!+ \sum_{j \in \Phi}\Gamma_{u_{j}} \mathcal{I}^{(u)}_{u}(\alpha) \sqrt{P_{u_j} h_j r_j^{-\eta}}+ \Gamma_s \sqrt{\beta P_b} \mathcal{I}^{(u)}_{s}(\alpha) + n_o,
\label{mathced_uplink}
\end{align}
\normalsize
where $*$ denotes the convolution operator, $S^{*}$ is the complex conjugate of $S$, and $n_o$ is a Gaussian random variable with zero mean and variance equals to $\sigma_{n}^2$ which is given by,
\begin{align}
    \sigma_{n}^2=\frac{N_o}{2} \int\nolimits_{-\frac{B_u+\alpha B}{2}}^{\frac{B_u+\alpha B}{2}} |H(f)|^2 df=\frac{N_o}{2},
\end{align}
and $\mathcal{I}^{(u)}_{s}$, $\mathcal{I}^{(u)}_{d}$ and $\mathcal{I}^{(u)}_{u}$ are the effective interference factors that depend on the amount of overlap between the uplink and downlink frequency bands (i.e., $\alpha$) and the used pulse shapes\footnote{In HD scheme, the interference factors capture the adjacent channel interference, which may occur due to the out-of-band ripples of $S_u(f)$ and $S_d(f)$. as shown in Fig. \ref{fig:EI}}. In general, $\mathcal{I}^{(a)}_{b} \in [0,1]$ represent the effective interference factor from $b$ on $a$, where $a$ and $b \in \{u,d\}$, and is given by
\begin{align}
\mathcal{I}^{(a)}_{b}(\alpha)& =\int\nolimits_{-\frac{B_a+\alpha B}{2}}^{\frac{B_a+\alpha B}{2}} S_b(f-f_b+f_a) S_a^{*}(f) df.
\label{equ:Idd}
\end{align}

\normalsize
Note that $\mathcal{I}^{(a)}_{s}(\alpha) = \mathcal{I}^{(a)}_{b}(\alpha)$ because SI is a form of cross mode interference.
To highlight the effect of the pulse shape and duplex parameter $\alpha$ on the effective interference factors $\mathcal{I}^{(u)}_d$ and $\mathcal{I}^{(d)}_u$ we plot Fig. \ref{fig:EI}. The figure shows that when the same pulse shape is used for the uplink and downlink, the effective interference factors are equivalent, however, this is not the general case when different pulse shapes are used in the uplink and downlink.

Since all UEs use the same pulse shape and the same frequency band, $\mathcal{I}^{(u)}_{u}$ is equal to unity. However, $\mathcal{I}^{(u)}_{d}$ represents the cross interference factor from the downlink on the uplink and is equal to $\mathcal{I}^{(u)}_{s}$ because SI is a cross-mode interference. Note that the partial overlap leads to  $\mathcal{I}^{(u)}_{s} < 1$  that can help to improve the self-interference cancellation.

Similar to \eqref{mathced_uplink}, the base-band downlink signal at the output of the matched filter of the test UE can be expressed as:
\vspace{-3.5pt}
\small
\begin{align}
&\!\!\!\!\!\!\!\!\!\!\!\!\!y_d(t_o)= A \sqrt{P_b r_o^{-\eta} h_o} + \sum_{k \in \Psi} \Gamma_{d_{k}} \mathcal{I}^{(d)}_{d}(\alpha) \sqrt{P_b h_k r_k^{-\eta}}  \notag \\
 &\!\!\!\!\!\!\!\!\!\!\!\!\!+ \sum_{j \in \Phi}\Gamma_{u_{j}} \mathcal{I}^{(d)}_{u}(\alpha) \sqrt{P_{u_j} h_j r_j^{-\eta}}+  \Gamma_{s} \sqrt{\beta P_{u_o}} \mathcal{I}^{(d)}_{s}(\alpha) + n_o,
 \label{mathced_downlink}
\end{align}\normalsize
where, $\mathcal{I}^{(d)}_{d}(\alpha)=1$, $\sigma_{n}^2=\frac{N_o}{2}$. $\mathcal{I}^{(d)}_{u}$ represent the cross interference factor from the uplink on the downlink and it has the same value as $\mathcal{I}^{(d)}_{s}$.

Inspecting \eqref{mathced_uplink} and \eqref{mathced_downlink} it is obvious that both $y_u(t_o)$ and $y_d(t_o)$ are conditionally (i.e., conditioning on the network geometry and channel gains) complex Gaussian random variable. As mentioned before, the conditional Gaussian representation of $y(t_o)$ allows us to calculate the conditional BER is the form of $ \omega_1 \text{erfc}\left( \sqrt{\omega_2\text{ SINR}} \right)$. Then, a consecutive averaging step over the network geometry is necessary to obtain the average BER. Now we calculate the SINR inside the erfc function, which is averaged over a Gaussian random interference-plus-noise. Let $\Xi_u=\{h_o, h_k , r_k,P_{u_j}, h_j,  r_j\}$, the conditionally average SINR for the uplink transmission can be expressed~as

\small
{\begin{align} \label{SINR_UL}
&  \!\!\!\!\!\!\!\!\!\!\!\!\! {\rm SINR}_u \left( \Xi_u\right) = \frac{| \mathbb{E} \left[ y_u(t_o)\big \vert  \Xi_u \right] |^2}{\text{Var}\left( y_u(t_o) \big \vert  \Xi_u  \right)}  \notag \\
&  \!\!\!\!\!\!\!\!\!\!\!\!\! = \frac{ {\rho h_o}}{\underset{{k \in {\Psi}}}{\sum}  P_b h_k r_k^{-\eta}|\mathcal{I}^{(u)}_{d}(\alpha)|^2 + \underset{j \in {\Phi}}{\sum} P_{u_j} h_j r_j^{-\eta}+\beta P_b |\mathcal{I}^{(u)}_{s}(\alpha)|^2+\sigma_n^2 } \notag \\
&  \!\!\!\!\!\!\!\!\!\!\!\!\! = \frac{h_o}{\underset{{k \in {\Psi}}}{\sum} \frac{P_{b} h_k r_{k}^{-\eta}|\mathcal{I}^{(u)}_d(\alpha)|^2}{\rho} + \underset{j \in {\Phi}}{\sum} \frac{{P_{u_j} h_j r_j^{-\eta}} }{{\rho}}   + \frac{{\beta} P_b | \mathcal{I}^{(u)}_s(\alpha)|^2}{{\rho}} + \frac{\sigma_n^2}{{\rho}}}.
\end{align}}
\normalsize

Similarly, let $ \Xi_d = \{h_o, r_o, h_k, r_k, P_{u_j}, P_{u_{o}},  h_j, r_j \}$,  the conditionally average (i.e., averaged over the Gaussian random variables) SINR for the downlink transmission is,

\small
{\begin{align}  \label{SINR_DL}
&  \!\!\!\!\!\!\!\!\!\!\!\!\! \!\!\!\!\!\! \!\!{\rm SINR}_d \left( \Xi_d\right) = \frac{|\mathbb{E} \left[ y_d(t_o)\big \vert  \Xi_d \right]|^2}{\text{Var}\left( y_d(t_o) \big \vert  \Xi_d  \right)}  \notag \\
&  \!\!\!\!\!\!\!\!\!\!\!\!\! \!\!\!\!\!\! \!\! = \frac{h_o}{\underset{{k \in \Psi}}{\sum} \frac{h_k r_{k}^{-\eta}}{ r_o^{-\eta}} + \underset{j \in \Phi}{\sum} \frac{{P_{u_j} h_j r_j^{-\eta}} |\mathcal{I}^{(d)}_u(\alpha)|^2}{{P_b r_o^{-\eta}}}+ \frac{{\beta} P_{u_o}  |\mathcal{I}^{(d)}_s(\alpha)|^2}{P_b  r_o^{-\eta}} + \frac{\sigma_n^2}{{P_b r_o^{-\eta}}}}.
\end{align}}

\normalsize
Now, $ \omega_1^{(u)} \text{erfc}(\sqrt{\omega_2^{(u)} {\rm SINR}_u(\Xi_u)})$ for the uplink and  $ \omega_1^{(d)} \text{erfc}(\sqrt{{\omega_2^{(d)} \rm SINR}_d(\Xi_d)})$ for the downlink, give the BER for a given realization of the cellular network at given time instant. Hence, we have to average over the random variables in $\Xi_u$ and $\Xi_d$. To do the averaging step, we exploit the following Lemma, which is given in \cite{Hamdi},

\footnotesize
\begin{align}
    \mathbb{E} \left[\omega_1 \text{erfc}\sqrt{\frac{\omega_2 x}{y+b}} \right]= \omega_1- \frac{\omega_1}{\sqrt{\pi} } \int\limits_{0}^{\infty} \frac{\mathcal{L}_y\left(\frac{z}{\omega_2}\right) e^{-z(1+\frac{b}{\omega_2})}}{\sqrt{z}}dz,
    \label{equ:BE}
\end{align}\normalsize
where $x$ is an exponential RV with unity mean, $y$ a non-negative RV with Laplace Transform (LT) $\mathcal{L}_y$ that is independent of $x$ and $b$ a constant. Projecting to the uplink case, we have $x =h_o$, $y=\sum_{k \in {\Psi}} \frac{P_{b} h_k r_{k}^{-\eta} |\mathcal{I}^{(u)}_d(\alpha)|^2}{\rho} + \sum_{j \in {\Phi}} \frac{{P_{u_j} h_j r_j^{-\eta}} }{{ \rho}}$ and $b=\frac{{\beta} P_b  | \mathcal{I}^{(u)}_s(\alpha)|^2}{{ \rho}} + \frac{\sigma_n^2}{{ \rho}}$. Let the LT of the RV variable y in this case be donated by $\mathcal{L}_{I_{u}}(z)$, then using \eqref{equ:BE}, the BER in the uplink is given by,

\footnotesize
\begin{align}
    &\textbf{BER}_{\mathcal{UL}}(\alpha)= \omega_1^{(u)}-\notag \\
     &\frac{\omega_1^{(u)}}{\sqrt{\pi} } \int\limits_{0}^{\infty} \frac{\mathcal{L}_{I_{u}}\left(\frac{z}{\omega_2^{(u)}}\right) e^{-z(1+\frac{{\beta} P_b  |\mathcal{I}^{(u)}_s(\alpha)|^2}{{\omega_2^{(u)} \rho}} + \frac{\sigma_n^2}{{\omega_2^{(u)} \rho}})}}{\sqrt{z}}dz.
    \label{equ:BEu}
\end{align}
\normalsize

Projecting \eqref{equ:BE} to the downlink case while conditioning on $r_o$, we have $x =h_o$, $y=\sum_{k \in \Psi} \frac{h_k r_{k}^{-\eta}}{r_o^{-\eta}}$ + $\sum_{j \in \Phi} \frac{{P_{u_j} h_j r_j^{-\eta}} |\mathcal{I}^{(d)}_u(\alpha)|^2}{{ P_b r_o^{-\eta}}}$, and $b= \frac{{\beta} P_{u_o} | \mathcal{I}^{(d)}_s(\alpha)|^2}{ P_b  r_o^{-\eta}} + \frac{\sigma_n^2}{ P_b r_o^{-\eta}}= \frac{{\beta} \rho |\mathcal{I}^{(d)}_s(\alpha)|^2}{ P_b  r_o^{-2\eta}} + \frac{\sigma_n^2}{ P_b r_o^{-\eta}}$.  Let the LT of the RV variable y in this case be donated by $\mathcal{L}_{I_{d}}(z)$, then using \eqref{equ:BE}, the BER in the downlink direction is given by,

\footnotesize
\begin{align}
    &\!\!\!\!\!\!\!\!\!\!\!\!\! \textbf{BER}_{\mathcal{DL}}(\alpha)= \omega_1^{(d)}-\notag \\
     &\!\!\!\!\!\!\!\!\!\!\!\!\! \mathbb{E}_{r_{o}} \left[\frac{\omega_1^{(d)}}{\sqrt{\pi} } \int\limits_{0}^{\infty} \frac{\mathcal{L}_{I_{d}}\left(\frac{z}{\omega_2^{(d)}}\right) e^{-z\left(1+ \frac{{\beta} \rho  |\mathcal{I}^{(d)}_s(\alpha)|^2}{\omega_2^{(d)} P_b  r_o^{-2\eta}} + \frac{\sigma_n^2}{{\omega_2^{(d)}  P_b r_o^{-\eta}}}\right)}}{\sqrt{z}}dz \right].
    \label{equ:BEd}
\end{align}

\normalsize
The BER expressions in both the uplink and downlink directions are characterized via the following theorem:

\begin{theorem}
\label{theorem1}
In a single-tier Poisson cellular network with channel inversion power control of threshold $\rho$, $\alpha B$ overlap between the uplink and downlink frequency bands, and exponentially distributed channel gains with unity means, the BER in the uplink and downlink directions for a generic user and a generic BS can be expressed as in equations (\ref{equ:BEuGen}) and (\ref{equ:BEdGen}),
where ${}_2 \text{F}_1 (.)$  is the Hypergeometric function, $\mathbb{E}_{P_u}\left[ \sqrt{P_u}\right]$, $\mathcal{I}^{(u)}_{d}(\alpha)$, $\mathcal{I}^{(d)}_{u}(\alpha)$, and $f_R(r_o)$ are given in \eqref{equ:powerD}, \eqref{equ:Idd}, and \eqref{equ:rDis}.
\rm Proof: see the Appendix.
\end{theorem}

\scriptsize
\begin{figure*}
\scriptsize
\begin{align}
\! \! \! \! \! \! \! \! \! \!  \textbf{BER}_{\mathcal{UL}}(\alpha)= \omega_1^{(u)}- \int\limits_{0}^{\infty}\frac{\omega_1^{(u)}}{\sqrt{\pi z}}  \exp \Bigg(&- \frac{2 \pi   \lambda}{ \eta-2} \frac{z}{\omega_2^{(u)} }\rho^{\frac{-2}{\eta}} \mathbb{E}_{P_u}\left[ P_u^{\frac{2}{\eta}}\right]
   {}_2 \text{F}_{1}\left(1,1-\frac{2}{\eta},2-\frac{2}{\eta},-\frac{z}{\omega_2^{(u)}}\right) - \frac{2}{\eta}\pi^{2} \lambda \left( \frac{z}{\omega_2^{(u)} \rho} P_b  |\mathcal{I}^{(u)}_{d}(\alpha)|^2\right)^{\frac{2}{\eta}}
     \text{csc} \left( \frac{2 \pi}{\eta} \right)\notag \\
        &-z\left(1+\frac{{\beta} P_b  |\mathcal{I}^{(u)}_s(\alpha)|^2}{{\omega_2^{(u)} \rho}} + \frac{\sigma_n^2}{\omega_2^{(u)} \rho}\right) \Bigg)dz.
    \label{equ:BEuGen}
\end{align}
\scriptsize \vspace{-0.15cm}
\begin{align}
\! \! \! \! \! \! \! \! \! \!      \textbf{BER}_{\mathcal{DL}}(\alpha)=
    \omega_1^{(d)}- \int\limits_{0}^{\infty} \int\limits_{0}^{\left(\frac{P^{(m)}_u}{\rho}\right)^{\frac{1}{\eta}}} \frac{\omega_1^{(d)} f_R(r_o)}{\sqrt{\pi z}}
     \exp \Bigg(&-z\left(1+ \frac{{\beta} \rho |\mathcal{I}^{(d)}_s(\alpha)|^2}{ \omega_2^{(d)} P_b  r_o^{-2\eta}} + \frac{\sigma_n^2}{{\omega_2^{(d)} P_b r_o^{-\eta}}}\right)
     - \frac{2\pi   \lambda z |\mathcal{I}^{(d)}_{u}(\alpha)|^2 {P_u}^{\frac{2}{\eta}} \rho^{1-\frac{2}{\eta}} r_o^{\eta} }{\omega_2^{(d)} (\eta-2) P_b }\notag \\
   &\! \! \! \! \! \! \! \! \! \! \! \! \! \! \! \! \! \! \! \! \! \! \! \!{}_2 \text{F}_1 \left(1,1-\frac{2}{\eta},2-\frac{2}{\eta},-\frac{z |\mathcal{I}^{(d)}_{u}(\alpha)|^2 \rho r_o^{\eta}}{ \omega_2^{(d)} P_b} \right)
    -\frac{2 \pi \lambda r_o^2 z}{(\eta-2)\omega_2^{(d)}} {}_2 \text{F}_1 \left(1,1-\frac{2}{\eta},2-\frac{2}{\eta},-\frac{z}{\omega_2^{(d)}} \right) \Bigg) dr_o dz.
    \label{equ:BEdGen}
\end{align}
\hrulefill
\end{figure*}

\normalsize

A particular case of interest is at $\eta = 4$, which is a practical value for outdoor communications for cellular network.  In this case, the BER  equations in \textbf{Theorem 1} reduce to equations \eqref{equ:BEu4} and \eqref{equ:BEd4}. 

\begin{figure*}
\scriptsize
\begin{align}
\! \! \! \! \! \! \! \! \! \! \! \! \textbf{BER}_{\mathcal{UL}}(\alpha) \overset{(\eta=4)}{=} \omega_1^{(u)}- \int_{0}^{\infty} \frac{\omega_1^{(u)}}{\sqrt{\pi z}}
  \exp\left( - \pi \lambda \sqrt{\frac{z}{\omega_2^{(u)} \rho}} \left(  \mathbb{E}_{P_u} \left[\sqrt{{P_u}}  \right]   \arctan\left(\sqrt{\frac{z}{\omega_2^{(u)}}}\right)
   + \frac{\pi }{2} \sqrt{P_b  |\mathcal{I}^{(u)}_{d}(\alpha)|^2}\right) -z \left(1+\frac{{\beta} P_b  \mathcal{I}^{(u)}_s(\alpha)}{\omega_2^{(u)} \rho} + \frac{\sigma_n^2}{\omega_2^{(u)} \rho} \right) \right) dz.
    \label{equ:BEu4}
\end{align}\vspace{-0.15cm}
\begin{align}
 \! \! \! \! \! \! \! \! \! \!     \textbf{BER}_{\mathcal{DL}}(\alpha) \overset{(\eta=4)}{=}
    \omega_1^{(d)}-\int\limits_{0}^{\infty} \int\limits_{0}^{\left(\frac{P^{(m)}_u}{\rho}\right)^{\frac{1}{4}}} \frac{\omega_1^{(d)} f_R(r)}{\sqrt{\pi z}}
      \exp \Bigg(&-z\left(1+ \frac{{\beta} \rho |\mathcal{I}^{(d)}_s(\alpha)|^2}{\omega_2^{(d)} P_b  r_o^{-8}} + \frac{\sigma_n^2}{{ \omega_2^{(d)} P_b r_o^{-4}}}\right)
     - \pi   \lambda \sqrt{\frac{z}{\omega_2^{(d)}}} r_o^2 \notag \\
      &\! \! \! \! \! \! \! \! \! \!  \Bigg( \sqrt{\frac{ |\mathcal{I}^{(d)}_{u}(\alpha)|^2}{P_b}}
    \mathbb{E}_{P_u}\left[ \sqrt{P_u}\right] \arctan \left(r_0^2 \sqrt{\frac{\rho z |\mathcal{I}^{(d)}_{u}(\alpha)|^2}{\omega_2^{(d)} P_b}} \right) + \arctan \left( \sqrt{\frac{z}{\omega_2^{(d)}}} \right) \Bigg) \Bigg) dr_o dz.
     \label{equ:BEd4}
\end{align}
\hrulefill
\end{figure*}
\normalsize
\section{Simulation Results} \label{Results}
In this section, we present numerical results and validate the proposed analysis with independent system-level simulations. In each simulation round, we generate a PPP cellular network over a $400 \text{Km}^2$ area, then we distribute the users randomly in the network until each BS has only 1 user who is able to maintain a threshold $\rho$. The collected results are taken for BSs and UEs located within $4 \text{Km}^2$ square at the center to avoid edge effects. At each transceiver in this region, we calculate the interference power due to other BSs and UEs at the matched filter end, then we evaluate the SINR for both uplink and downlink from \eqref{SINR_UL} and \eqref{SINR_DL}, respectively. Unless otherwise stated, we use the parameters values listed in Table 1, and for the pulse shapes, we employ rectangular and triangular pulses in the time, which reduce  to ${\rm sinc(.)}$ and ${\rm sinc}^2(.)$ in the frequency domain.
\begin{table} []
\caption{\; Simulation parameters.}
\centering
\begin{tabular}{|l|l|l|l|}
\hline
\rowcolor[HTML]{C0C0C0}
\textbf{Parameter} & \textbf{Value}      & \textbf{Parameter} & \textbf{Value} \\ \hline
$\rho$             & -70 dBm             & $P_b$              & 5 W            \\ \hline
$\lambda$          & 3 $\text{BSs/Km}^2$ & $P^{(m)}_u$          & 1 W            \\ \hline
$B_u$              & 1 MHz                & $B_d$              & 1 MHz           \\ \hline
$\beta$            & -80 dB                   & $N_o$              & -90 dBm
           \\ \hline
$\omega_1^{(u)}$            & 1                   & $\omega_2^{(u)} $             & 1              \\ \hline
$\omega_1^{(d)}$            & 1                   & $ \omega_2^{(d)}$             & 1              \\ \hline
\end{tabular}

\end{table}

\begin{figure}[t]
\centerline{\includegraphics[width=  2.75in]{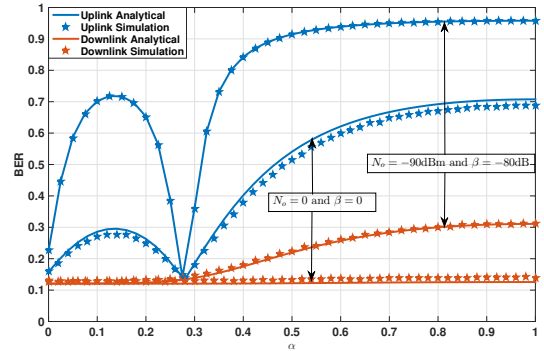}}\caption{\,: BER vs $\alpha$ simulation and analytical.}
\label{fig:BERSim}
\end{figure}

\begin{figure}[t]
\centerline{\includegraphics[width=  2.75in]{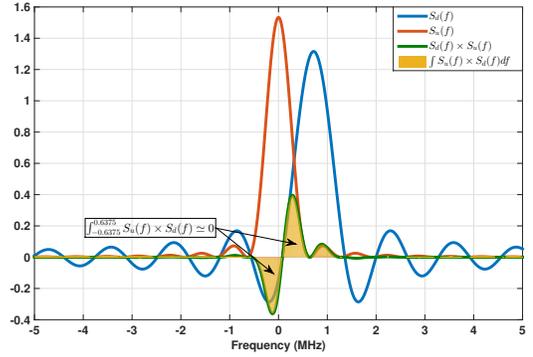}}\caption{\,: The FT of the rectangular (downlink) and triangular (uplink) pulse shapes at the optimal balanced $\alpha=0.275$.}
\label{fig:MinAlpha}
\end{figure}

First, we study the BER performance against $\alpha$ to show the gradual effect of the uplink/downlink spectrum overlap, with and without SI and noise effects, on the BER of both the uplink and downlink. As observed in Fig.  \ref{fig:BERSim}, the effect on the uplink/downlink overlap is not monotone on the BER, thanks to the pulse shaping. In addition, the cost of cross mode interference is more prominent on the uplink than on the downlink. With perfect SI cancellation (i.e., $\beta=0$), the BER in the downlink is almost unaffected, while the uplink BER increased with up to $350 \%$ at the FD case (i.e., $\alpha=1$). Accounting for residual SI, the BER degradation is more significant on both the uplink and downlink especially at the FD case.  Interestingly, we also find that there is a range of spectrum overlap that can provide more BW for the uplink and downlink to access without significant cost in the BER. The main conclusion from the figure is that operating at the FD (i.e., $\alpha =1$) may not be a practical solution due to the uplink vulnerability to cross mode interference. It is worth highlighting that the proposed $\alpha$-duplex scheme may provide an improved BER than the HD scheme  due to the adjacent channel interference caused by the out-of-band ripples of $S_u(f)$ and $S_d(f)$ as will shown in Fig.~\ref{fig:MinAlpha}.

\begin{figure}[t]
\centerline{\includegraphics[width=  2.75in]{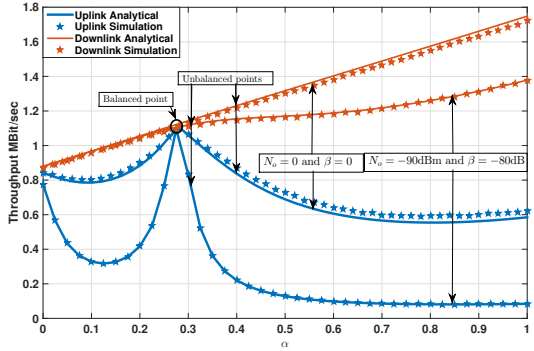}}\caption{\,: Throughput vs $\alpha$ simulation and analytical.}
\label{fig:ThrSim}
\end{figure}

In the second simulation example, we investigate the advantage of the BW improvement by studying the throughput of the same links in the previous example vs $\alpha$ in Fig. \ref{fig:ThrSim}. Similar to the BER performance trend,  the improvement in the downlink throughput, obtained at full spectrum overlap, comes on the expense of a degradation in the uplink throughput. On the other hand, the benefit of increasing the BW while achieving approximately the same BER is obviously observed at specific $\alpha$ ranges. Hence, FD may not be the best duplexing scheme that maximizes the throughput for the uplink and downlink  in cellular networks, specially with imperfect SI cancellation. Therefore, the proposed $\alpha$-duplex scheme provides a fine tuned overlap between uplink and downlink spectrum which can be optimized to maximize the throughput. For instance, if a symmetric uplink and downlink connectivity is required, $\alpha \approx 0.275$ is the optimal operating point, denoted as {\em balanced point} in Fig. \ref{fig:ThrSim}. In contrast, if more data rate is required in the downlink, then $\alpha$ need to be optimized to maximize the downlink throughput subject to an acceptable  degradation in the uplink throughput. For instance, the point denoted by {\em unbalanced point} on Fig. \ref{fig:ThrSim} maximizes the downlink throughput subject to no degradation in the uplink throughput.

To elaborate the existence of optimal overlap of the uplink and downlink, we plot the FT of the rectangular pulse shape (for downlink) and the triangular pulse shape (for uplink) at the optimal balanced $\alpha$ in Fig. \ref{fig:MinAlpha}. The figure shows that at a certain overlap point between the uplink and downlink the pulse shapes are almost orthogonal which makes the effective interference factors $I^{(u)}_d$ and $I^{(d)}_u$ close to zero. Therefore, at this point of interest, the BSs and UEs have a larger bandwidth to access without the cost of cross-mode interference. The figure also shows the out-of-band ripples of $S_d(f)$ and $S_u(f)$ that cause adjacent channel interference in the case of HD operation as shown in Fig. \ref{fig:EI}, which leads to the superiority of the $\alpha$-duplex in the BER over the HD scheme.

\begin{figure}[t]
\centerline{\includegraphics[width=  2.75in]{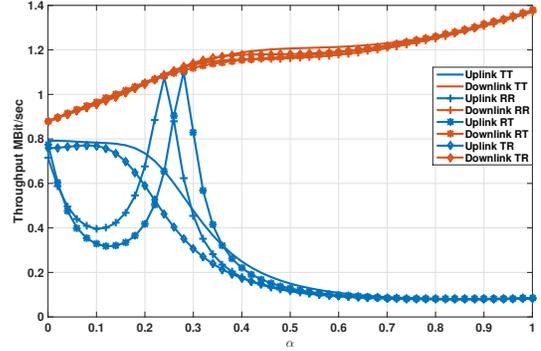}}\caption{\,: Throughput vs $\alpha$ using different pulse shapes.
}
\label{fig:Throughput}
\end{figure}

Finally, we study the impact of adopting different types of pulse shapes  on the throughput performance in Fig. \ref{fig:Throughput}, and on finding the optimal spectrum overlap. In the absence of SI, cross-mode interference has a negligible effect on the downlink. Hence, in this case, the pulse shapes do not affect the  downlink throughput. On the other hand, the choice of the pulse shapes drastically affects the throughput in the uplink direction. Therefore, pulse shaping can be exploited to move the optimal overlap to higher values in order to increase the spectral efficiency and the gains harvested via SI cancellation. Fig. \ref{fig:Throughput} clearly shows the superiority of the proposed $\alpha$-duplex scheme over the FD scheme in cellular networks. In contrast to the FD scheme, which {\em improves} the downlink throughput by $55.6 \%$ on the cost of {\em degrading} the uplink throughput by $87.5 \%$ with-respect-to the HD scheme, the proposed $\alpha$-duplex scheme, under balanced network operation and $\mathrm{RT}$ pulse shapes provides $22.5\%$ and $37.5\%$ {\em improvement} in the downlink and uplink throughput with respect to the HD scheme,~respectively.


\section{Conclusion}\label{Conclusion}

A tractable framework for in-band FD communications is developed. The model is used to shed the light on the vulnerability of uplink to downlink interference and argue that full overlap between uplink and downlink channels may not be a practical solution in cellular networks. We propose a fine tuned partial overlap between uplink and downlink frequencies, denoted as $\alpha$-duplex scheme, to optimize the overall network performance. To this end, the results show that there exists an optimal value for the overlap parameter $\alpha$ which depends on the network parameter and design objective. Finally, the impact of pulse shaping and SI cancellation is investigated through simulation results.

\appendix

First we need to find the LT of the aggregated interference from both sources; BSs and UEs. Let the LT of the term $\sum_{k \in {\Psi}} \frac{P_{b} h_k r_{k}^{-\eta} |\mathcal{I}^{(u)}_d(\alpha)|^2}{\rho}$ be donated  by $\mathcal{L}_{\mathcal{I}_{d \rightarrow u}}(s)$ and the LT of $\sum_{j \in {\Phi}} \frac{{P_{u_j} h_j r_j^{-\eta}} }{{\rho}}$ be donated by $\mathcal{L}_{\mathcal{I}_{u \rightarrow u}}(s)$. Then  $\mathcal{L}_{\mathcal{I}_{u}}(s) = \mathcal{L}_{\mathcal{I}_{d \rightarrow u}}(s) \mathcal{L}_{\mathcal{I}_{u \rightarrow u}}(s)$. The LT $\mathcal{L}_{\mathcal{I}_{d \rightarrow u}}(s)$ can be expressed as

\scriptsize
 \begin{align}
  & \mathcal{L}_{\mathcal{I}_{d \rightarrow u}}(s) = \mathbb{E}\left[e^{-  \frac{s}{\rho}\underset{x_i \in {\mathbf{\Psi}}\setminus \{o\}}{\sum}  P_{b,i} |\mathcal{I}^{(u)}_{d}(\alpha)|^2 h_i \left\|x_i\right\|^{-\eta}}\right]\notag \\
  &\stackrel{(i)}{=}  \mathbb{E}_{\mathbf{\Psi}}\left[\underset{x_i \in {\mathbf{\Psi}}\setminus \{o\}}{\prod} \mathbb{E}_{h_i}\left[e^{-\frac{s}{\rho} P_{b,i}  |\mathcal{I}^{(u)}_{d}(\alpha)|^2 h_i \left\|x_i\right\|^{-\eta}}\right]\right]\notag \\
  &\stackrel{(ii)}{=}  \exp\left( - 2 \pi   \lambda \int_{0}^{\infty} \mathbb{E}_{h}\left[  \left(1- e^{-\frac{s}{\rho} P_b  |\mathcal{I}^{(u)}_{d}(\alpha)|^2 h x^{-\eta}}\right)\right] xdx \right)\notag \\
   &\stackrel{(iii)}{=} \exp\left( - \frac{2}{\eta}\pi^{2} \lambda \left( \frac{s}{\rho}  P_b  |\mathcal{I}^{(u)}_{d}(\alpha)|^2\right)^{\frac{2}{\eta}} \text{csc} \left( \frac{2 \pi}{\eta} \right) \right),
\end{align}\normalsize
where $\mathbb{E}_x[.]$ is the expectation with respect to the random variable $x$. $(i)$ follows from the independence between $\mathbf{\Psi}$ and $h_i$. $(ii)$ using the probability generation functional (PGFL) of PPP. $(iii)$ using the LT of $h$.

Following the same steps but accounting for the interference protection region defined by$ \left(P_{u,i} \left\|x_i\right\|^{-\eta} > \rho \right)$ around each BS as in \cite{uplink_h}, $ \mathcal{L}_{\mathcal{I}_{u \rightarrow  u}}(s)$ is obtained as


\scriptsize
\begin{align}
 &\!\!\!\!\!\!\!\!\!\!\!\!\!\!\!\! \mathcal{L}_{\mathcal{I}_{u \rightarrow  u}}(s) =\mathbb{E}\left[e^{-  \frac{s}{\rho} \underset{x_i \in {\mathbf{\Phi}}\setminus \{o\}}{\sum} \mathbbm{1}\left(P_{u,i} \left\|x_i\right\|^{-\eta} < \rho \right) P_{u,i} h_i \left\|x_i\right\|^{-\eta}}\right]\notag \\
  &\!\!\!\!\!\!\!\!\!\!\!\!\!\!\!\!=  \exp \left( - 2 \pi   \lambda s \rho^{\frac{-2}{\eta}} \mathbb{E}_{P_u}\left[ P_u^{\frac{2}{\eta}}\right] \int_{s^\frac{-1}{\eta}}^{\infty} \frac{y}{1+y^{\eta}} dy \right)\notag \\
  &\!\!\!\!\!\!\!\!\!\!\!\!\!\!\!\!=  \exp \left( - \frac{2 \pi   \lambda}{\eta-2} s \rho^{\frac{-2}{\eta}} \mathbb{E}_{P_u}\left[ P_u^{\frac{2}{\eta}}\right] {}_2 \text{F}_{1}\left(1,1-\frac{2}{\eta},2-\frac{2}{\eta},-s\right) \right).
  \end{align}

\normalsize
Hence, the overall LT of the interference that affects the uplink transmission is given by,

\scriptsize
\begin{align}
 & \!\!\!\!\!\!\!\!\!\!\!\!\!\!\!\!\!\! \mathcal{L}_{\mathcal{I}_{u}}(s) =\exp \left( - \frac{2 \pi   \lambda}{\eta-2} s \rho^{\frac{-2}{\eta}} \mathbb{E}_{P_u}\left[ P_u^{\frac{2}{\eta}}\right] {}_2 \text{F}_{1}\left(1,1-\frac{2}{\eta},2-\frac{2}{\eta},-s\right)  \right.\notag \\
  & \left.\quad \quad \quad \quad \quad \quad \quad \quad - \frac{2}{\eta}\pi^{2} \lambda \left( \frac{s}{\rho} P_b  |\mathcal{I}^{(u)}_{d}(\alpha)|^2\right)^{\frac{2}{\eta}} \text{csc} \left( \frac{2 \pi}{\eta} \right) \right).
  \label{equ:LSu}
\end{align}
\normalsize
Substituting (\ref{equ:LSu}) in (\ref{equ:BEu}) we get  (\ref{equ:BEuGen}).

For the downlink direction, following similar steps, let the LT of the term $\sum_{k \in \Psi} \frac{h_k r_{k}^{-\eta}}{ r_o^{-\eta}}$ be donated  by $\mathcal{L}_{\mathcal{I}_{d \rightarrow d}}(s)$ and the LT of $\sum_{j \in \Phi} \frac{{P_{u_j} h_j r_j^{-\eta}} |\mathcal{I}^{(d)}_u(\alpha)|^2}{{P_b r_o^{-\eta}}}$ be donated by $\mathcal{L}_{\mathcal{I}_{u \rightarrow d}}(s)$. Then the needed LT  $\mathcal{L}_{\mathcal{I}_{d}}(s)$ is found by the product of these two values.
For $\mathcal{L}_{\mathcal{I}_{d \rightarrow d}}(s)$, the interference protection region is defined by $\left\|x_i\right\|> r_o$ due to the closest BS associations, following the same steps as before,

\scriptsize
\begin{align}
  & \mathcal{L}_{\mathcal{I}_{d \rightarrow d}}(s|r0) =\exp\left( - 2 \pi   \lambda \int_{r_o}^{\infty}\mathbb{E}_{h}\left[  \left(1- e^{- s h r_o^{\eta} x^{-\eta}}\right)\right] xdx \right)\notag \\
 &{=}  \exp\left(\frac{-2 \pi \lambda r_o^2 s}{\eta-2} {}_2 \text{F}_1 \left(1,1-\frac{2}{\eta},2-\frac{2}{\eta},-s \right)\right).
\end{align}\normalsize
For $\mathcal{L}_{\mathcal{I}_{u \rightarrow d}}(s|r0)$, we approximate the location of the tagged UE to be the same as his serving BSs location (collocated) as \cite{Que}. Based on this approximation, $\mathcal{L}_{\mathcal{I}_{u \rightarrow d}}(s|r0)$ is given by

\scriptsize
\begin{align}
  & \mathcal{L}_{\mathcal{I}_{u \rightarrow d}}(s|r0){=} \notag \\
 &\exp\left( - 2 \pi   \lambda \int_{(\frac{P_u}{\rho})^{\frac{1}{\eta}}}^{\infty}\mathbb{E}_{P_u,h}\left[  \left(1- e^{-\frac{s P_u  |\mathcal{I}^{(d)}_{u}(\alpha)|^2 h x^{-\eta}}{P_b r_o^{-\eta}}}\right)\right] xdx \right)\notag.
 \end{align}
 \begin{align}
   &=  \exp\Bigg( - \frac{2\pi   \lambda s |\mathcal{I}^{(d)}_{u}(\alpha)|^2 \mathbb{E}_{h}\left[ {P_u}^{\frac{2}{\eta}} \right] \rho^{1-\frac{2}{\eta}} r_o^{\eta} }{(\eta-2) P_b } \notag \\
   &\quad \quad \quad \quad \quad \quad \quad \quad{}_2 \text{F}_1 \left(1,1-\frac{2}{\eta},2-\frac{2}{\eta},-\frac{s |\mathcal{I}^{(d)}_{u}(\alpha)|^2 \rho r_o^{\eta}}{P_b} \right) \Bigg).
\end{align}
\normalsize
Hence, the LT of the overall interference affecting the downlink is given by:
\scriptsize
\begin{align}
& \!\!\!\!\!\!\!\!\!   \mathcal{L}_{\mathcal{I}_{d}}(s|r0) =  \exp \Bigg( - \frac{2 \pi \lambda s}{\eta-2} \left( r_o^2 {}_2 \text{F}_1 \left(1,1-\frac{2}{\eta},2-\frac{2}{\eta},-s \right) \right. +  \notag \\
   &    \frac{|\mathcal{I}^{(d)}_{u}(\alpha)|^2 \mathbb{E}_{h}\left[ {P_u}^{\frac{2}{\eta}} \right] r_o^{\eta} }{  \rho^{\frac{2}{\eta}-1} P_b } {}_2 \text{F}_1 \left(1,1-\frac{2}{\eta},2-\frac{2}{\eta},-\frac{s |\mathcal{I}^{(d)}_{u}(\alpha)|^2 \rho r_o^{\eta}}{P_b} \right) \Bigg).
 \label{equ:LSd}
\end{align}

\normalsize
Substituting (\ref{equ:LSd}) in (\ref{equ:BEd}) we get  (\ref{equ:BEdGen}).

\bibliographystyle{IEEEtran}
\bibliography{ref}

\vfill

\end{document}